\documentclass[acmsmall]{acmart}
\AtBeginDocument{%
  \providecommand\BibTeX{{%
    \normalfont B\kern-0.5em{\scshape i\kern-0.25em b}\kern-0.8em\TeX}}}

\renewcommand\footnotetextcopyrightpermission[1]{}

\usepackage{fancyhdr}
\usepackage{subcaption}

\usepackage{booktabs}
\usepackage{tabularx}
\usepackage{longtable}
\usepackage{multirow,bigstrut}
\usepackage{multirow}
\usepackage[linesnumbered,ruled,vlined]{algorithm2e}
\begin{document}

\title{Manuscript of a method for improving wear in intermittently computing file systems}

\author{Yeteng Liao}
\authornote{Both authors contributed equally to this research.}
\email{}
\orcid{0000-0002-6372-8133}
\author{Han wang}
\authornotemark[1]
\email{willaaaaaawhh@gmail.com}
\affiliation{%
  \institution{East China Normal University}
  \streetaddress{}
  \city{}
  \state{}
  \country{China}
  \postcode{}
}

% \affiliation{%


\begin{abstract}
  For the first time, the repeated wear phenomenon of high-frequency power failure on the data block area in intermittent computing file system is found. A method to improve NVM wear in ICFS under high-frequency power failure scenarios is proposed.
\end{abstract}





\maketitle

\section{Introduction}
%With the development of ultra-low power processing technology and wireless communication technology, micro-sensing devices can be widely connected to the Internet. Often referred to as the Internet of Things (IoT), such applications can be instantiated as sensor networks, healthcare implants and wearables%
As embedded systems and software technologies continue to advance, embedded devices are increasingly used in various applications, including IoT devices, wearable devices, and healthcare implants. Sensor devices that are one kind of IoT device have a wide range of applications, such as environmental sensing \cite{b1}, remote sensing \cite{b2}, and wildlife tracking and monitoring \cite{b3}. 
However, batteries' size limitation and high maintenance cost make it challenging to embed them in some scenarios. In such cases, battery-free energy-harvesting devices that can harvest energy from sources such as radio waves, light, or wind have emerged as a solution. These sustainable and clean energy sources allow energy harvesting devices to operate for extended periods without requiring battery replacements or maintenance.

Environmental energy harvesting is inherently unstable, as it can be affected by various factors such as weather conditions. 
This instability can lead to a lower energy level, which can result in a loss of computational results.  
To address this issue, energy harvesting devices utilize checkpoints to store the state of the program in persistent memory.   
This allows the device to resume operation from the last checkpoint when energy is restored, ensuring that no data is lost due to energy depletion\cite{intro-ck}\cite{Achieving}\cite{Mementos}.

By incorporating large capacitors on energy harvesting devices, their capabilities can become increasingly advanced, including low-power Bluetooth communication\cite{BLE}, machine learning, and intelligent perception\cite{stateful-neural}\cite{Intelligence}.
It can also offer services for the emerging field of edge computing\cite{Edge-Computing}, such as improving data processing and storage capabilities in these environments. 
As a device processes more data, its local storage requirements increase. This makes a reliable file system crucial for intermittent computing. With the emergence of the first ICFS, iNVMFS, new possibilities have opened up for intermittent computing. It uses lower power consumption and lower latency NVM as the storage medium.

Popular types of NVM, such as ReRAM, FeRAM, STT-RAM, PCM, have a smaller write tolerance compared to SRAM/DRAM(about 1000 times lower), as stated in \cite{Emerging-NVM}. 
When the number of writes to the unit exceeds the persistence threshold, the unit loses the ability to retain data without applying power, which is known as persistence. Since the main benefit of using NVM devices is their persistence, repeatedly writing the same cells can be problematic.

iNVMFS\cite{iNVMFS} organize file data into blocks for management while ensuring the consistency of the program state with the stored state. However, this approach also presents a challenge: an unstable energy supply will always lose  post-checkpoint data and need to be re-written. This can cause blocks of data to be written repeatedly. This is because, as mentioned in [8], [11], and [12], the most common file operations in sensor applications involve appending new data to existing files, rather than updating existing file data. When appending data to files and a power failure occurs, the data must be rewritten after recovery. Under high-frequency power failures, this can result in repeated re-writes, leading to significant wear and tear on the FRAM and reducing its service life.

In this study, we introduce a wear-leveling method that is suitable for the first NVM- based  intermittently computational file system. Our study makes two key contributions:
\begin{itemize}
\item Firstly, we explore the impact of checkpoint frequency and outage frequency on the wear and tear of NVM-based ICFS. This is the first study and provides important insights into how the frequency of checkpoints and power failure affect the durability and longevity of NVM in ICFS.

\item Secondly, We propose a high-frequency power failure detection method based on ICFS implementation.

\item Finally, we propose a wear improvement and leveling method method that is specifically designed for NVM-based ICFS. This method takes into account the findings of our exploration and distributes write operations in a way that minimizes wear on the NVM, which includes a dynamic buffer tuning strategy, a write operation migration strategy, and a write balancing strategy.
\end{itemize}
Overall, our study offers important insights into the design and optimization of NVM-based ICFS, and provides a practical solution for reducing wear and improving the durability of this type of system.
\section{Background and Motivation}

\subsection{Background}

iNVMFS is the first file system on an interruptible computing device where checkpoints of storage space are managed by iNVMFS when checkpoints of program state are committed. Both metadata and user data are stored in FRAM. Reads of metadata are redirected to the log area through a hash table, and changes to metadata are recorded in the log space, which ensures metadata consistency. 
Changes to user data are directly performed in FRAM, and the consistency of user data is ensured by the in-place overwriting technique.   For the modification of metadata recorded in the log area, when the checkpoint comes, a $commit$ record will be recorded by the file system in the log area, and any data before the commit is considered as the data before the checkpoint and write back to the metadata area, so as to achieve the consistency of the program state and storage state.  When a power failure occurs,  the log space cursor goes back to the $commit$ record, ensuring that the metadata is consistent with the program state.

\begin{figure}[h]
    \centering
    \includegraphics[width=0.90\linewidth]{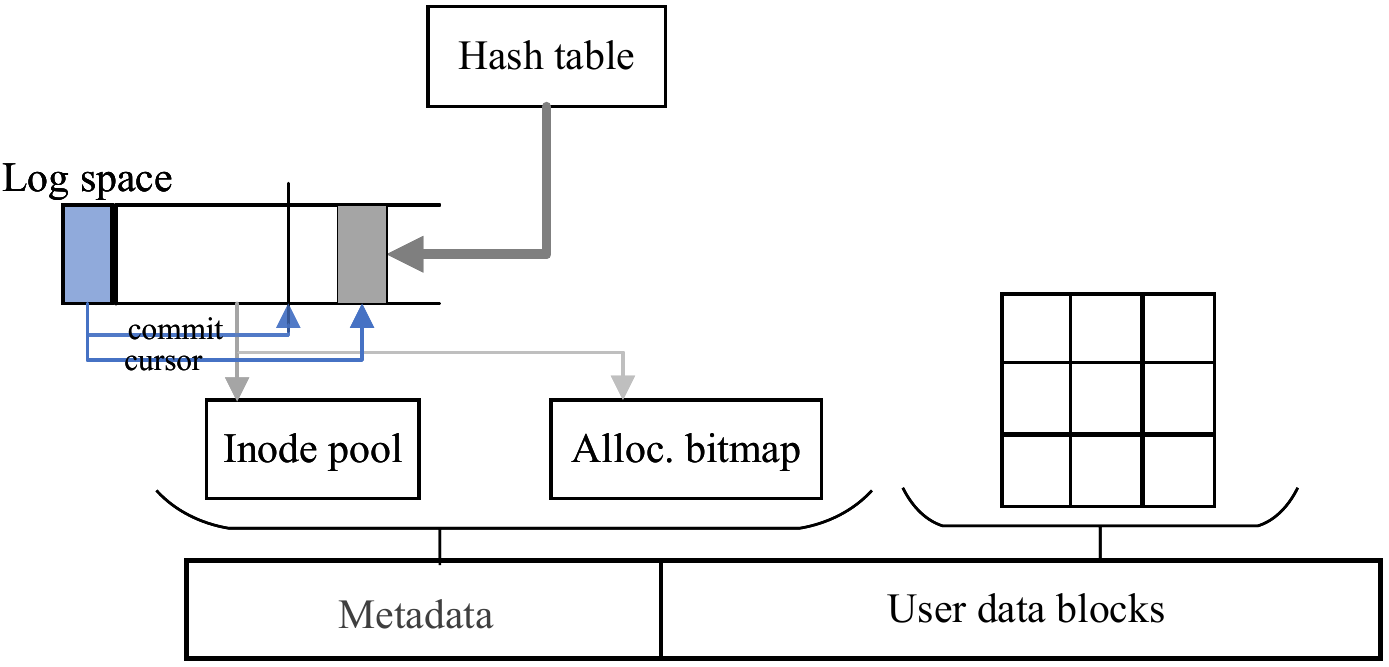}
    \caption{description1 }
    \label{fig:background}
\end{figure}

\subsection{Related Works}

There has been significant research in intermittent computing systems, such as the works referenced in \cite{Adaptive-checkpoint}\cite{Achieving}\cite{Efficient-ck}\cite{Fast-CK}\cite{Efficient-state}\cite{Fixing-ck}, which employ checkpointing mechanisms to save program state snapshots to enable the program to resume from where it left off after a power outage. While these methods improve program progress, there is still some loss of progress post-checkpoint.

The JIT mechanism saves a snapshot of the program state to non-volatile memory when a power failure is imminent \cite{Mementos}\cite{supporting}\cite{hibernus++}. The availability of this snapshot allows intermittent computing systems to resume without experiencing the loss of progress post-checkpoint. However, there are some challenges with the JIT approach. For example, the requirement for a voltage monitor limits its applicability. The JIT mechanism requires an ADC (Analog to Digital Converter) to detect an imminent power failure, as noted in \cite{Mementos} and \cite{Failure-sentinels}. This additional hardware requirement increases the cost and applicability of interruptible computing systems. Furthermore, the JIT mechanism generates an order of magnitude higher overhead than the checkpointing mechanism, as reported in \cite{Forget-failure}, \cite{Failure-sentinels}, \cite{Adaptive-checkpoint}, and \cite{Achieving}. The design of this paper is independent of program state checkpointing techniques.

Wear balancing in NVM file systems has been extensively researched in the field of non-intermittent computing, with various techniques proposed to mitigate wear on inode data (which experiences the majority of wear in the field of non-intermittent computing). For example, MARCH\cite{MARCH} uses a window-based indirect pointer implementation strategy to collect write operations into a moving window and distribute them into PCM space by sliding windows. VInode\cite{VInode} breaks the binding between logical inodes and inode slots by virtualizing inode tables and balances inode slot wear by migrating inodes with higher write loads. Themis\cite{Themis} protects NVM from malicious write operations by migrating inodes and data pages originally stored in NVM to volatile memory. LMWM\cite{LMWM} strikes a balance between wear accuracy, data migration, and metadata management overhead by switching data migration granularity between 64 bytes and page size (e.g., 4 KB). Contour\cite{Contour} uses inode virtualization to decouple inodes from physical memory addresses, enabling them to move to different memory locations for wear leveling.

However, these techniques are not applicable to intermittent computing systems that are energy unstable due to the possibility of introducing new inconsistency problems. In addition to this, such systems suffer from frequent power failures which cause severe wear on the data block area, something that is not focused on in non-intermittent computing. Therefore, new techniques are needed to address the unique challenges posed by intermittent computing systems.

\subsection{Motivation}
\begin{figure}[h]
    \centering
    \includegraphics[width=0.8\linewidth]{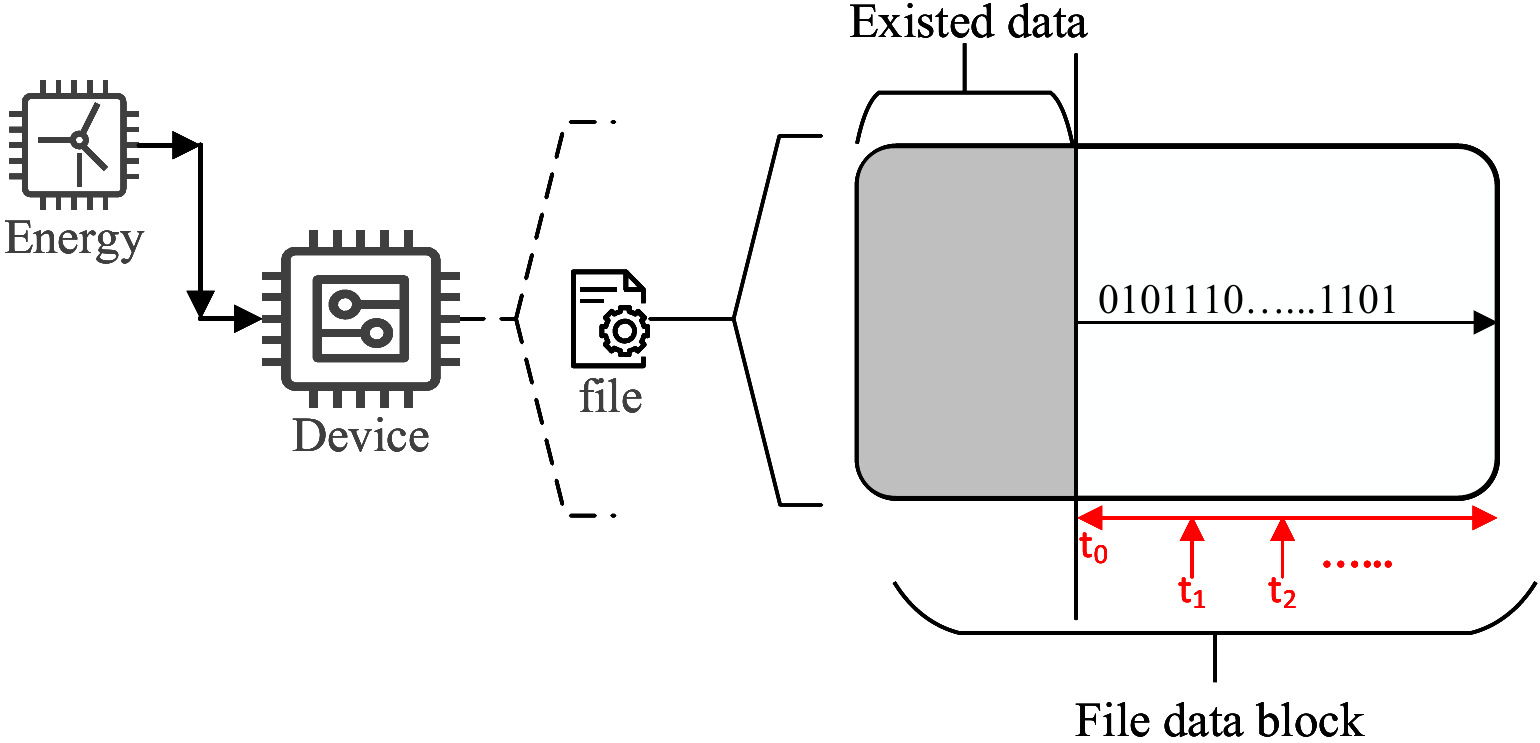}
    \caption{description }
    \label{fig:motivation_1}
\end{figure}

High-frequency checkpoints can lead to frequent backup operations, which can result in a significant energy overhead. However, if the interval between checkpoints is too large, a program that takes a long time to execute may never be complete, as it may lose all progress made before the next checkpoint.
In a reasonable checkpoint setup, the ICFS might have the following conditions for post-checkpoint data:

In this study, ICFS (intermittent computing file system) stores user data in FeRAM, just like the first ICFS \textemdash iNVMFS. 
For devices with intermittent computing, energy instability is always present. Especially in the field or in bad weather conditions, high-frequency power failure events will occur frequently. 
Fig. \ref{fig:motivation_1} shows the problems of today’s ICFS under high-frequency power failure.

Under a high-frequency power failure, the sensor application appends data to a data block of a file. 
As shown in Fig. \ref{fig:motivation_1}, the writing starts at the moment $t_0$. When it reaches moment $t_1$, the power failure occurs. 
After the power failure is recovered, the program state and the stored state of the program are restored to the state at the moment $t_0$. 
Then, the execution continues forward. When writing reaches moment $t_2$, the power failure occurs again,  and after recovery again, it continues to write from time $t_0$. 
With a high-frequency power failure, this situation will be repeated numerous times.
The high frequency of power failure not only causes many invalid writes but also causes severe wear on the data block, which greatly reduces the lifetime of NVM.

Of course, a high frequency of checkpointing can alleviate this problem, but a high frequency of checkpointing leads to frequent backup operations, and frequent backups not only consume limited energy but also cause causes program interruptions, which brings high overhead\cite{Adaptive-checkpoint}.
\subsubsection{Observation One}
\begin{figure}[h]
    \centering
    \includegraphics[width=1\linewidth]{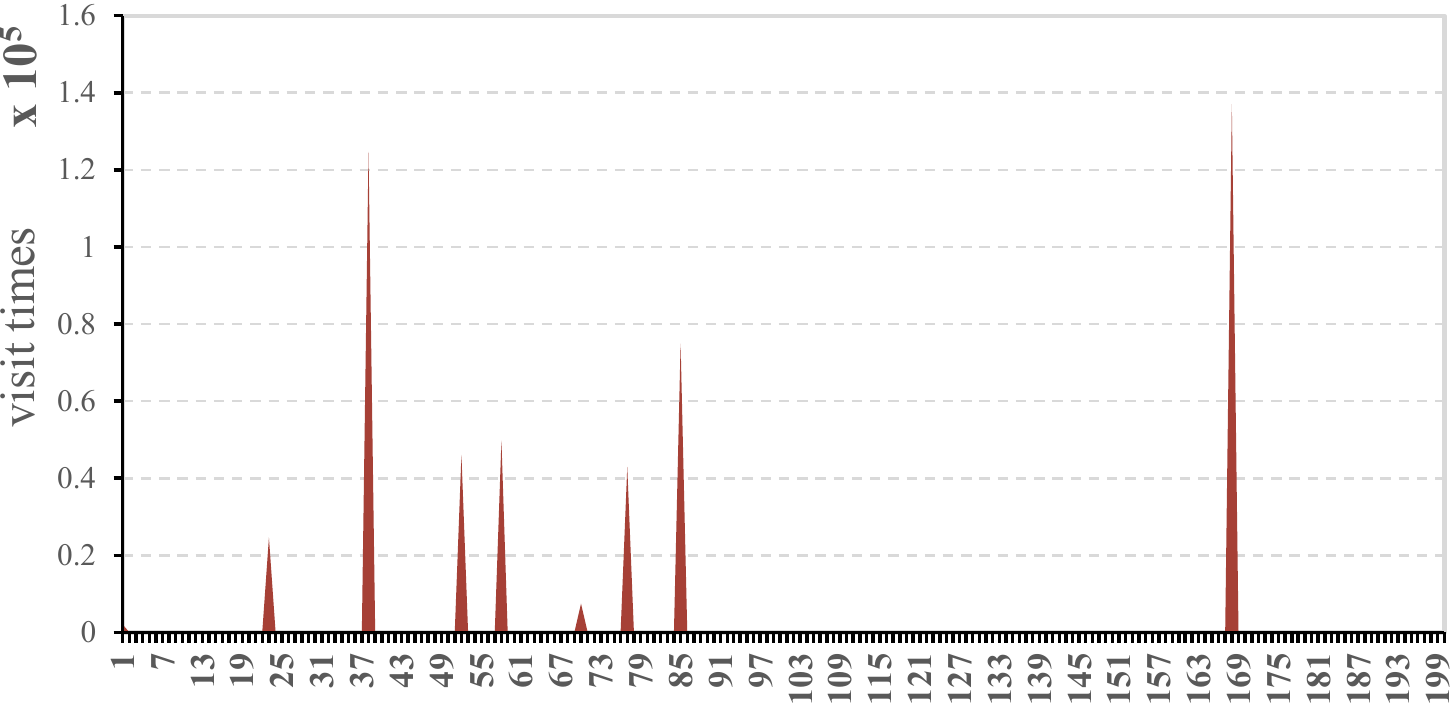}
    \caption{Write distribution on iNVMFS }
    \label{fig:observation_1}
\end{figure}

When power failures occur frequently (with a power failure frequency of 0.6 and a checkpoint frequency of 10), the number of visits to each data block can be seen in Fig. \ref{fig:observation_1}. 
It's evident that the number of visits to several data blocks has exceeded $10^5$, causing severe wear and tear on the NVM. This significantly reduces the NVM's service life and increases maintenance costs.

\subsubsection{Observation Two}
%用的最大值，跑了2000次的平均最大值
We set the checkpoint interval to an average of every 10 writes as a checkpoint, with 16B writes at once and a $20\%$ probability of power failure, (the amount of data that needs to be written is 4KB). 
Through the experiment, we observe that under the high frequency of power failure, a lot of wear is caused between the two checkpoints, and the maximum number of writes is 138. 
This means that 2208B of data is written between the two checkpoints. 
The amount of data written to the two checkpoints should have been 160B, which increased the amount of data written by $1280\%$.
This increases the number of writes by $1280\%$, greatly reducing the lifetime of this memory cell. 
Of course, the situation becomes even worse if the checkpoints are set too far apart.

\section{Design}
\subsection{Overview}

\begin{figure}[h]
    \centering
    \includegraphics[width=1\linewidth]{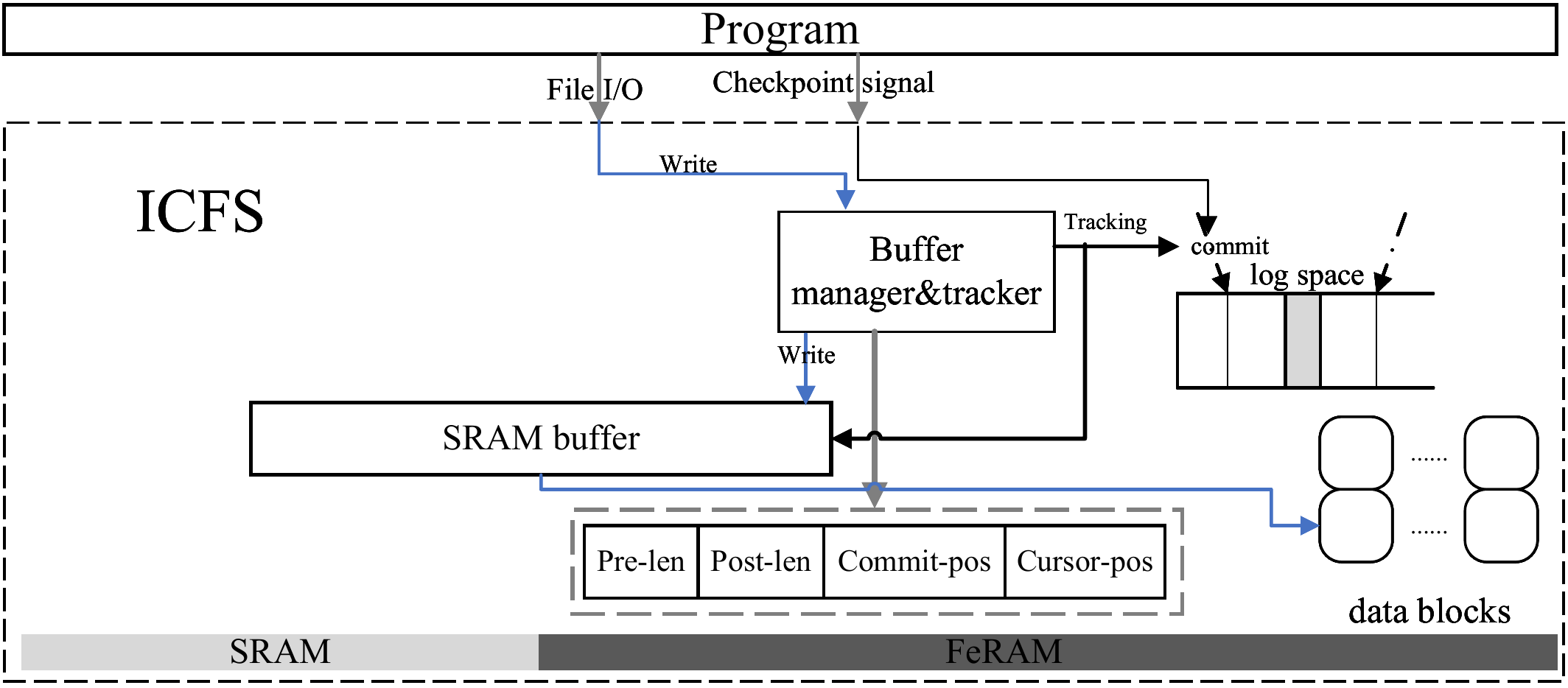}
    \caption{Design }
    \label{fig:overview}
\end{figure}

Fig. \ref{fig:overview} provides an overview of the proposed wear-leveling design for an ICFS. The ICFS includes three areas, namely the user data area, the log area, and the metadata area, which are all located in non-volatile memory (NVM). To reduce user block wear, we have implemented a wear-leveling strategy that offloads frequent writes to user block regions to an SRAM buffer. Moreover, to minimize the space usage of SRAM, we design the buffer size to be dynamic. Additionally, we have implemented an adaptive approach that migrates writes to the buffer only when a high-frequency power failure is detected, thus minimizing the energy overhead.

\subsection{Metadata management}
To be improved
\subsection{Wear improvement }

\subsubsection{power failure detection}

High-frequency power failure detection technology

We have designed a lightweight high-frequency power outage detection technique that selectively enables buffer mode only in cases of high-frequency power outages. This approach prevents excessive wear and tear on the Non-Volatile Memory (NVM) during frequent power outages and allows for data to be immediately written back when energy becomes available. For a detailed overview of our approach, please consult Algorithm \ref{A1}. We use the $Cursor_{pos}$ and $Commit_{pos}$ positions in the ICFS (iNVMFS) log space to decide whether to move a write operation to the buffer. 
When we receive the checkpoint signal $C_{sig}$, we update the $commit$ position $Commit_{pos}$. Each time the $C_{sig}$ signal arrives, it is considered a success, and for each success, the number of failures $FailCount$ is set to zero. 
Each time the position of $Cursor{-pos}$ is the same as $Commit{-pos}$ (we use the iNVMFS log space to roll back the cursor to the commit point after a power outage is recovered), we count it as a failure. For each failure, $SuccessCount$ is set to zero. 
When the number of $FailCount$ exceeds the failure threshold $FailThreshold$, the buffer is turned on, and the number of $SuccessCount$ exceeds the success threshold $SuccessThreshold$, the buffer is turned off.

\begin{algorithm}[H]
    \SetAlgoLined %显示end
	\caption{Buffer Management Algorithm with Signal }\label{A1}%算法名字
	\KwIn{$Commit{-pos}$ , $Cursor{-pos}$ , $C_{sig}$ }\ %输入参数 
	% \KwOut{output result}%输出
	% some description\; %\;用于换行
        % 初始化
		% $r_{len} \leftarrow -1$;\tcp{Initialization}\
        \tcp{Initialization}\
        $FailCount \gets 0$;
        $SuccessCount \gets 0$;
        $BufferStatus \gets$ "Inactive"\;
        \While{true}{
            \If{$C_{sig}$ is received}{
                $SuccessCount \gets SuccessCount + 1$\;
                $FailCount \gets 0$\;
                \If{$SuccessCount \geq SuccessThreshold$}{
                    $BufferStatus \gets$ "Inactive"\;
                    % $FailCount \gets 0$\;
                }
            }
            \If{$Cursor{-pos} == Commit{-pos}$}{
                $FailCount \gets FailCount + 1$\;
                $SuccessCount \gets 0$\;
                \If{$FailCount \geq FailThreshold$}{
                    $BufferStatus \gets$ "Active"\;
                    % $FailCount \gets 0$\;
                }
            }
            
        }
        % \If{$C_{sig}$ is received}{
        %     $Commit{-pos} \gets Cursor{-pos}$; %需要记录cursor的位置吗？？
        % }
        return
\end{algorithm}

\subsubsection{Write reduction}
%%要改 ！！！%
When ICFS needs to append data to a file block, it redirects the write operation to a buffer and atomically writes back to the block when the checkpoint signal is received. To accomplish this, ICFS maintains a data block access information table $V_{info}$. 
See Algorithm \ref{A2} for details:
in the event of a high-frequency power failure, ICFS offloads the write operation to a buffer in the SRAM. First, the address of the data block to be written is written to the buffer, followed by the actual data (lines 18-20). When the checkpoint signal $C_{sig}$ is received, ICFS atomically writes the data back to the NVM data area (lines 23) based on the address stored in the buffer.
If the amount of data to be written $data_w$ exceeds the current block capacity $curr_{size}$, ICFS needs to find an unallocated block with the minimum access times $V_b$ based on the access information table $V_{info}$ for the next write. The value of $data_w$ is then set to 0 (lines 25-30).

\begin{algorithm}[H]
    \SetAlgoLined %显示end
	\caption{algorithm write}\label{A2}%算法名字
	\KwIn{ $V_{info}$, $B_{ad}$, $C_{sig}$,$curr_{size}$,$data_{w}$,$V_{b}$,$r_{len}$,$Pre_{len}$}\ %输入参数 curr_{size} 可以写入的大小 V_b块的访问次数
	% \KwOut{output result}%输出
	% some description\; %\;用于换行
        % $B_{ad}$ 写回地址
        \If{not high-frequency power failure}{
            $r_{len} \leftarrow -1$, $Pre_{len} \leftarrow -1$;\tcp{Initialization}\
            \For{ $i$ in $C_{sig}$}{  % 信号点到来  
                $filelen_i \leftarrow track()$\;
                \eIf{$Pre_{len}$ == -1}{
                    $Pre_{len} \leftarrow filelen_i$\; 
                } 
                {
                    $r_{len} \leftarrow (filelen_i - Pre_{len})$\;
                    $Pre_{len} \leftarrow filelen_i$\;
                    
                }
                \If{log space is written back}{
                    $Pre_{len} \leftarrow -1$\;
                }
		  }
        }
        \Else{
                %BufferStatus
                \If{$BufferStatus$ is "Active"}{
                    $buffer \gets malloc(r_{len})$\;
                    $buffer \leftarrow$ $T_{ad}$\;
    			$buffer \leftarrow data$\;
                }
                % 信号点到来 根据地址，原子性写回到数据块中
                \If{$C_{signal}$}{
                    $T_{ad}$ $\stackrel{atomically}\longleftarrow$ data\;
                }
                \If{$data_{w}$ $>$ $curr_{size}$}{ % 已写入的数据大于当前容量了 换块
    		      \If{$V_{b}$ is minimum and unallocated}{
                        change block\;
                        $data_{w} \leftarrow 0$\;
                    }
                }
		}
        return 
\end{algorithm}

\subsubsection{Dynamic buffer size}
To optimize the use of SRAM space, we dynamically adjust the size of the buffer based on the growth of data. We achieve this by analyzing the file system records in the log space and checkpoint commit records to understand the data growth pattern. This allows us to minimize wasted SRAM space during power outages. When a checkpoint signal is detected (which the iNVMFS can detect), we search the log space for changes in file size within the metadata. By calculating the increase in file size between two checkpoints, we can estimate the amount of data to be written, and adjust the buffer size accordingly with low overhead.
% 有问题%
The details of this process are outlined in Algorithm \ref{A3}. We store the value of $r_{len}$ in the metadata area of ICFS and set the buffer size based on this value. We initialize $r_{len}$ to -1 (line 1) and monitor the checkpoint signal $C_{sig}$. When $C_{sig}$ arrives, we find the length of the file $filelen_i$ (lines 2-3) stored in the metadata through the hash table in the original system. If $r_{len}$ is still -1, we assign the currently updated file length directly to $r_{len}$ (lines 4-5). Otherwise, we calculate the change in file length (lines 6-9) by subtracting the length of the last record from the current file size. We repeat this process several times to estimate the amount of data to be written between the two checkpoints, and dynamically adjust the buffer size accordingly.
Finally, when the log space performs a write-back operation, we initialize $r_{len}$ to -1 again (lines 10-12). By dynamically adjusting the buffer size according to the data growth pattern, we can optimize the use of SRAM space and minimize waste.

\begin{algorithm}[H]
    \SetAlgoLined %显示end
	\caption{algorithm Dynamic buffer size}\label{A3}%算法名字
	\KwIn{$r_{len}$,$C_{sig}$ }\ %输入参数 
	% \KwOut{output result}%输出
	% some description\; %\;用于换行
        % 初始化
		$r_{len} \leftarrow -1$;\tcp{Initialization}\
        \For{ $i$ in $C_{sig}$}{  % 信号点到来  
            $filelen_i \leftarrow track()$\;
            \eIf{$r_{len}$ == -1}{
                $r_{len} \leftarrow filelen_i$\; 
            } 
            {
                $last_{len} \leftarrow r_{len}$\;
                $r_{len} \leftarrow (filelen_i - last_{len})$\;
            }
            \If{log space is written back}{
                $r_{len} \leftarrow -1$\;
            }
		}
        return
\end{algorithm}

\subsection{Wear-leveling improvement}
Find the block with the minimum number of accesses, and maintain a visit information table.

\section{Evaluation}

\subsection{Setup}
SRAM has 8KB space and FeRAM has 256KB space in our experiments. 
The storage space allocated to the file system is 128KB and the space allocated to the user data block is 100KB, each user data block is 512B (same as the iNVMFS setting). 
In the 100KB storage space, we write some file data, which takes up about 50KB. 
we perform an append write operation to a file (the sensor application has almost no random writes and update operations) and write 4KB data. 

% on average, one checkpoint is set for every ten append operations, and 16B of data is written atomically for each append operation. 
% The power failure rate is set to 20\%.
\begin{table}[htbp]
  \centering
  \caption{Add caption}
    \begin{tabular}{cccc}
    \toprule
    Parameters & \multicolumn{3}{c}{Descriptions} \\
    \midrule
    one append operation & \multicolumn{3}{c}{atomic 16B} \\
    checkpoint frequency & \multicolumn{3}{c}{One checkpoint for every 10 append operations.} \\
    power failure rate & \multicolumn{3}{c}{0.2} \\
    \bottomrule
    \end{tabular}%
  \label{tab:setup}%
\end{table}%

To evaluate the proposed techniques, several related methods are compared in this paper as follows: 
\begin{itemize}
    \item BL is the baseline method that randomly selects an  unallocated block for writing as used in \cite{iNVMFS}.
    \item  TP is the naive method based on threshold setting. 
    When the number of writes to a data block reaches a threshold(The default is set to 30), the block is exchanged Randomly prioritize blocks with 0 visits, or blocks that are less than the threshold and unallocated, and finally select unallocated blocks.
    \item TM is the method based on TP, but it always selects the new block with the least number of writes for new arrivals.
    \item  \textbf{BF} is the proposed technique in this paper, which take advantage of a dynamic buffer in SRAM to reduce the writes to blocks, and combines the greedy-based block selection strategy.
\end{itemize}

We use the average number of writes, $\mu$, and standard deviation, $\sigma$, to evaluate the performance of wear-leveling. As shown in Equation (1) and (2), respectively, where $n$ is the total number of blocks, $x_i$ is the number of writes of the $i$th block.
Then, to evaluate the degree of block fragmentation, a metric, $F$, is used, the calculation is shown in Equation (3), where $U$ is the number of used data blocks, and $T$ is the total number of data blocks. 
% This metric is similar to other works for evaluating the degree of memory fragmentation and storage fragmentation \cite{}\cite{}.
\begin{equation}
    \mu = \frac{\sum_{i=1}^{n}x_i}{n}
\end{equation}
\begin{equation}
  \sigma = \sqrt{\frac{\sum_{i=1}^{n}(x_i-\mu)^2}{n}}
\end{equation}
\begin{equation}
  F = 1- \frac{U}{T}
\end{equation}
A smaller value of the standard deviation indicates that the write operations performed on the block are relatively uniform, while a larger value suggests the opposite. A smaller value of F indicates a higher level of fragmentation, while a larger value of F indicates less fragmentation.

\subsection{Performance}

\subsubsection{Wear leveling}
Fig. \ref{fig:performance} (a) (b) shows the degree of wear and tear of data blocks in FeRAM-based ICFS.
It is noticeable that BF has a significant decrease compared to BL, with a 77\% decrease in standard deviation and a 69.50\% decrease in average visit times.
In comparison to BL, TP, and TM have a standard deviation reduction of 64.87\% and 67.36\%, respectively, and a 1.19\% decrease in average access times.

\subsubsection{Block fragmentation}
\begin{figure}[h]
    \centering
    \includegraphics[width=1\linewidth]{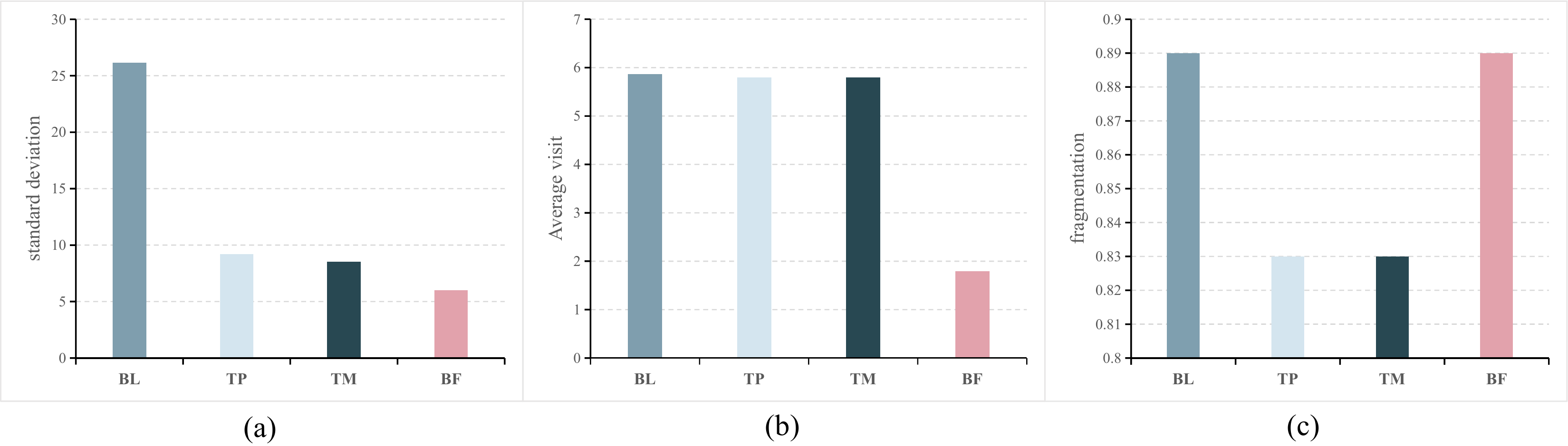}
    \caption{PFR = 0.2,CF = 10,threshold = 30 }
    \label{fig:performance}
\end{figure}

Fig. \ref{fig:performance}(c) shows the fragmentation generated by the four methods, and it is clear that our proposed method does not generate additional fragmentation. However, TP and TM perform poorly, and they generate an additional 6.74\% of wasted space.
% In order to measure the degree of wearing on the data block, we use standard deviation which is used as a statistical indicator of how spread out the data is. 

% \begin{equation}
%   \sigma = \sqrt{\frac{\sum_{i=1}^{n}(x_i-\mu)^2}{n}}
% \end{equation}
% where $\sigma$ denotes the standard deviation, $n$ denotes the total number of data blocks, $x_i$ denotes the number of write operations on the $i$th data block, and $\mu$ denotes  the average of write operations on all data blocks.

% We use equation (2) to calculate the fragmentation rate:
% \begin{equation}
%   F = 1- \frac{U}{T}
% \end{equation}
% where F denotes the fragmentation rate, U denotes the number of used data blocks, and T denotes the total number of data blocks. A smaller value of F indicates a higher level of fragmentation, while a larger value of F indicates less fragmentation.

\subsection{Sensitive study}
We conducted experiments for different power failure rates and checkpoint frequencies. The experimental results are shown in Table \ref{tab:sensitive}

% Table generated by Excel2LaTeX from sheet 'Sheet2'
\begin{table}[htbp]
  % \centering
  \caption{Add caption2}
    \begin{tabular}{c|c|c|c|c|c|c|c|c|c|c|c|c|c|}
    \toprule
    \multicolumn{1}{r}{} &       & \multicolumn{3}{c|}{CF = 5} & \multicolumn{3}{c|}{CF = 10} & \multicolumn{3}{c|}{CF = 15} & \multicolumn{3}{c|}{CF = 20} \\
    \midrule
    \multicolumn{1}{c}{PFR} &       & $\sigma$    & $\mu$ & F     & $\sigma$    & $\mu$ & F     & $\sigma$  & $\mu$ & F & $\sigma$  & $\mu$ & $F$ \\
    \midrule
    \multirow{4}[7]{*}{0.2} & BL    & 12.89 & 3.24  & 0.89  & 26.15  & 5.87  & 0.89  & 59.87  & 12.41 & 0.89  & 147.53 & 27.19 & 0.89 \\
\cmidrule{2-14}          & TP    & 8.01  & 3.19  & 0.85  & 9.21  & 5.8   & 0.83  & N/A   & N/A   & N/A   & N/A   & N/A   & N/A \\
\cmidrule{2-14}          & TM    & 7.95  & 3.19  & 0.85  & 8.53  & 5.8   & 0.83  & N/A   & N/A   & N/A   & N/A   & N/A   & N/A \\
\cmidrule{2-14}          & BF    & 6.01  & 1.79  & 0.89  & 6.01  & 1.79  & 0.89  & 6.01  & 1.79  & 0.89  & 6.01  & 1.79  & 0.89 \\
\midrule
    \multicolumn{1}{r}{} &       & \multicolumn{3}{c|}{CF = 5} & \multicolumn{3}{c|}{CF = 10} & \multicolumn{3}{c|}{CF = 15} & \multicolumn{3}{c|}{CF = 20} \\
    \midrule
    \multicolumn{1}{c}{PFR} &       & $\sigma$    & $\mu$ & F     & $\sigma$    & $\mu$ & F     & $\sigma$  & $\mu$ & F & $\sigma$  & $\mu$ & $F$ \\
    \midrule
    \multirow{4}[7]{*}{0.3} & BL    & 20.67 & 4.83  & 0.89  & 73.98    & 15.1  & 0.89  & 308.96   & 59.29 & 0.89  & 1508.91  & 257.45 & 0.89 \\
\cmidrule{2-14}          & TP    & 9.15  & 4.77  & 0.82  & N/A   & N/A   & N/A   & N/A   & N/A   & N/A   & N/A   & N/A   & N/A \\
\cmidrule{2-14}          & TM    & 8.92  & 4.77  & 0.82  & N/A   & N/A   & N/A   & N/A   & N/A   & N/A   & N/A   & N/A   & N/A \\
\cmidrule{2-14}          & BF    & 6.01  & 1.79  & 0.89  & 6.01  & 1.79  & 0.89  & 6.01  & 1.79  & 0.89  & 6.01  & 1.79  & 0.89 \\
\midrule
    \multicolumn{1}{r}{} &       & \multicolumn{3}{c|}{CF = 5} & \multicolumn{3}{c|}{CF = 10} & \multicolumn{3}{c|}{CF = 15} & \multicolumn{3}{c|}{CF = 20} \\
    \midrule
    \multicolumn{1}{c}{PFR} &       & $\sigma$    & $\mu$ & F     & $\sigma$    & $\mu$ & F     & $\sigma$  & $\mu$ & F & $\sigma$  & $\mu$ & $F$ \\ 
    \midrule
    \multirow{4}[7]{*}{0.4} & BL    & 37.08 & 8.15  & 0.89  & 271.68   & 52.69 & 0.89  & 2437.55  & 451.32 & 0.89  & HD    & HD    & HD \\
\cmidrule{2-14}          & TP    & 10.52 & 8.1   & 0.77  & N/A   & N/A   & N/A   & N/A   & N/A   & N/A   & N/A   & N/A   & N/A \\
\cmidrule{2-14}          & TM    & 10.08 & 8.1   & 0.77  & N/A   & N/A   & N/A   & N/A   & N/A   & N/A   & N/A   & N/A   & N/A \\
\cmidrule{2-14}          & BF    & 6.01  & 1.79  & 0.89  & 6.01  & 1.79  & 0.89  & 6.01  & 1.79  & 0.89  & HD    & HD    & HD \\
\bottomrule
    \end{tabular}%
  \label{tab:sensitive}%
\end{table}%

Observations can be made that:
\begin{itemize}
\item As the checkpoint and power failure frequencies increase, data blocks in ICFS are subject to severe wear and tear.
\item In cases where power failure frequencies are high and checkpoint frequencies are set unreasonably, it is difficult for the program to execute to completion. This observation is consistent with the conclusions of some research \cite{performance-analysis}.
\item While TP and TM can reduce wear and tear in certain situations, they are not available at slightly higher power failure and checkpoint frequencies. This is because they cause all data blocks in the file system to be allocated, leaving no unallocated blocks available, resulting in a great waste of already limited space.
\item The method proposed in this paper not only greatly reduces data block wear, but also does not waste additional space. It is not affected by checkpoint settings and high-frequency power failure.
\end{itemize}

\begin{figure}[h]
    \centering
    \includegraphics[width=1\linewidth]{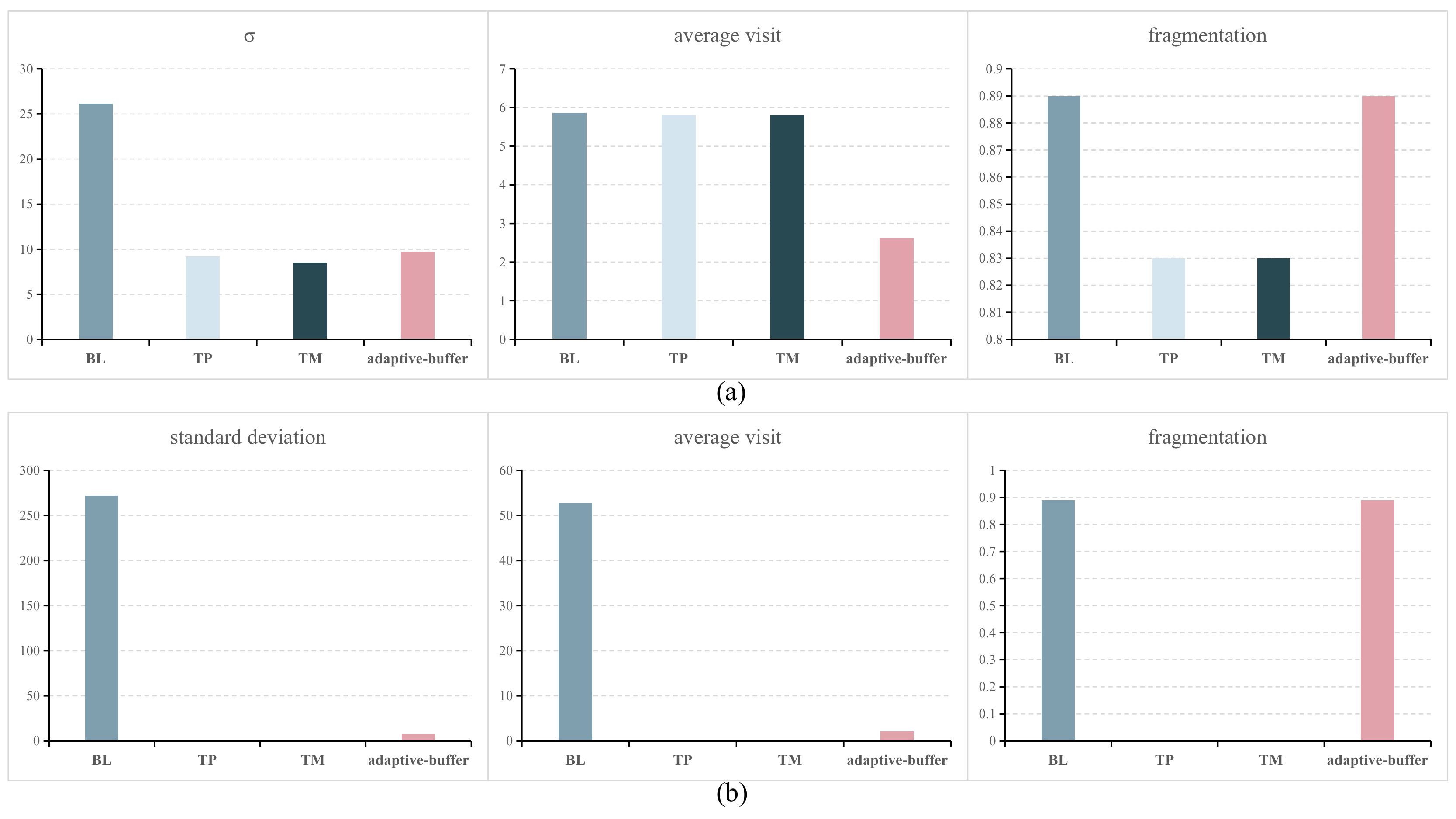}
     \caption{adaptive buffer, (a) 0.2 power failure rate,(b) 0.4 power failure rate }
    \label{fig:adaptive-buffer}
\end{figure}

\subsection{Overhead analysis}

1. Space

2. Time

\bibliographystyle{ACM-Reference-Format}
\bibliography{bibfile}

\end{document}